# Marginalised Poisson Hurdle Model for Cross-Sectional Count Data with Excess Zeros

Fred Fosu Agyarko[1,2*], Edward Acheampong[1,4,5], Issah Seidu[1], Samuel Iddi[1,3*]

[1]Department of Statistics and Actuarial Science, University of Ghana, Legon, Accra, Ghana

[2]Council for Scientific and Industrial Research (CSIR), Institute for Scientific and Technological Information (INSTI), Accra, Ghana

[3]Data Synergy and Evaluation Unit, African Population and Health Research Center, Nairobi, Kenya

[4]Department of Mathematics, Saveetha School of Engineering, SIMATS, Saveetha University, India

[5]School of Mathematical Sciences, University of Nottingham, Nottingham, United Kingdom

[*]**Corresponding authors**: Fred Fosu Agyarko (agyarkofosu@gmail.com), Samuel Iddi (siddi@ug.edu.gh)

## Abstract

**Background**: Count data with excess zeros arise frequently in health economics and epidemiology. The standard Poisson Hurdle Model (PHM) parametrises the underlying Poisson rate λ directly, so its count-component coefficients are log-rate ratios rather than log-ratios of the overall marginal mean $E(Y) = \mu$. As a result, the incidence density ratio (IDR) from the PHM is neither exact nor constant across covariate values, which complicates reporting in applied research.

**Methods**: We propose the Marginalised Poisson Hurdle Model (MPHM), which reparametrises the count component so that $\boldsymbol{\beta}$ directly governs $\log E(Y_i) = \mathbf{x}_i^\top \boldsymbol{\beta}, i = 1,2,\ldots,n$. A nonlinear connector equation links the structural Poisson rate $\lambda_i$ to the directly parametrised marginal mean $\mu_i$. We prove existence and uniqueness of the connector solution, develop a vectorised numerical connector solver (Brent's method), derive the exact score equations and block-diagonal Fisher information, establish asymptotic normality, and prove that $IDR_j = \exp(\beta_j), j = 1,2,\ldots,p$ is exactly constant across all covariate values.

**Results**: A simulation study with n ∈ {100, 250, 500, 1000}, π ∈ {0.2, 0.4, 0.6, 0.8}, and R = 200 replications confirms consistency, near-zero bias, and correct 95% Wald coverage (0.905–0.975) across all 16 scenarios. Applied to the National Medical Expenditure Survey 1988 (NMES1988) physician visit data (*n* = 4,406), the MPHM yields IDR = 1.163 (95% CI: 1.150–1.177) for each additional chronic condition, an exact, constant, population-wide effect not derivable from the standard PHM.

**Conclusions**: The MPHM resolves the non-constant IDR problem in Poisson hurdle models by directly parametrizing, $E(Y)$. The resulting IDR, $\exp(\beta_j)$, is exact and portable, holding for every individual and for the entire population without requiring further marginalisation. The model is straightforward to implement via Broyden–Fletcher–Goldfarb–Shanno (BFGS) maximum likelihood with a vectorised numerical connector solver, and its improved interpretability substantially simplifies reporting of covariate effects in health utilisation research.

## 1. Introduction

Count data with excess zeros are encountered routinely in health economics, epidemiology, and public health research. In many health surveys, a large proportion of participants report zero service utilisations, not because the count distribution naturally generates zeros, but because structural barriers such as lack of insurance, geographic distance, or financial constraint prevent access altogether. In the National Medical Expenditure Survey (NMES1988) analysed in this paper, 23.6% of individuals report zero physician office visits, compared with only 2.1% predicted by a fitted Poisson model. This substantial gap reflects a genuine dual process: some individuals face barriers that prevent any care-seeking, while others simply had no medical need during the survey period.

The Poisson Hurdle Model (PHM)[1] is the most widely used tool for this type of data in health research. Its appeal lies in a natural clinical interpretation: the hurdle separates the binary access decision (whether to seek care at all) from the intensity decision (how many visits, conditional on seeking care). The two components can have different covariate sets, can be estimated separately through log-likelihood factorisation, and the model handles both zero-inflation and zero-deflation. For these reasons, the PHM is often preferred over the zero-inflated Poisson[2] in health utilisation applications (Min and Agresti[3]; Deb and Trivedi[4]).

Despite its widespread use, the standard PHM has a fundamental limitation that receives too little attention in applied work. Because the count component parametrises $\log(\lambda_i)$ rather than $\log E(Y_i)$ for $i = 1,2,\ldots,n$, the regression coefficients are log-rate ratios, not log-mean ratios. The overall marginal mean, $E(Y_i) = \frac{(1-\pi_i)\cdot\lambda_i}{(1-e^{-\lambda_i})}$ is a nonlinear function of $\lambda_i$, and the IDR for covariate $\mathbf{x}_j$ for $j = 1,2,\ldots,p$ therefore varies across individuals depending on their covariate profile. In practice, this means that a single reported IDR from the PHM is not truly portable: it describes a specific hypothetical individual, not all individuals equally.

$$E(Y_i) = (1-\pi_i)\cdot\frac{\lambda_i}{1-e^{-\lambda_i}}, \tag{1}$$

which is a nonlinear function of $\lambda_i$. Consequently, the ratio $\frac{E(Y|x_j+1)}{E(Y|x_j)}$, the incidence density ratio (IDR) for covariate $x_j$ , is not equal to $\exp(\gamma_j)$ in the PHM, it depends on the current values of $\lambda_i$ and $\pi_i$, which vary across individuals. Reporting $\exp(\hat{\gamma}_j)$ as an IDR from the PHM is therefore not strictly justified and can yield misleading conclusions.

The natural solution is to parametrise $E(Y_i)$ directly. By replacing the PHM count component $\log(\lambda_i) = \mathbf{x}_i^\top\boldsymbol{\gamma}$ with $\log(\mu_i) = \mathbf{x}_i^\top\boldsymbol{\beta}$ where $\mu_i = E(Y_i)$, so that $\exp(\beta_j) = \frac{E(Y|x_j+1)}{E(Y|x_j)}$ becomes an exact constant IDR. This marginalisation approach was pioneered by Heagerty[8] for binary outcomes and extended to count data with excess zeros by Long et al.[7] for zero-inflated Poisson models. However, to the best of our knowledge, no marginalised counterpart of the Poisson hurdle model has been proposed in the literature. The present paper addresses this gap.

The MPHM introduces a nonlinear connector equation that links $\lambda_i$ implicitly to $\mu_i$ and $\pi_i$. We show that this equation always has a unique positive solution, derive a complete theoretical framework including identification, score equations, block-diagonal Fisher information, asymptotic normality, and a formal proof that $\exp(\beta_j)$ is an exact, constant IDR. We also show that the commonly used linear approximation $\lambda \approx \frac{\mu}{1-\pi}$ always overestimates the true solution and produces inconsistent estimators when substituted into the likelihood.

Our contributions are fourfold. First, we propose the MPHM as the first cross-sectional marginalised Poisson hurdle model and establish its theoretical properties through five formal propositions. Second, we demonstrate that the linear approximation of the connector equation is invalid and that the exact numerical solution is essential for reliable inference. Third, we conduct a simulation study across 16 factorial scenarios confirming consistency, near-zero bias, and correct coverage. Fourth, we apply the MPHM to the NMES1988 physician visit data (n = 4,406), illustrating the practical interpretational advantage through IDR comparison tables, Vuong tests, and goodness-of-fit analyses.

Section 2 develops the proposed analytical framework for MPHM, beginning with a detailed description of the data, followed by the standard PHM approach. It introduces the key connector equation that links the model components, establishes its theoretical properties, and derives the formal constant IDR result. Section 3 reports the simulation study. Section 4 applies the proposed MPHM to the NMES1988 data set. Section 5 discusses the limitations of the study and concludes the paper.

## 2. Methods

### 2.1 Case study

We utilize the 1987–1988 National Medical Expenditure Survey (NMES1988) as the motivating dataset for the study. The dataset ($n$ = 4,406) is available in the R package AER [10] and contains physician office visits from the US civilian non-institutionalised population. The outcome variable is the number of physician office visits. Covariates include: *health* (self-reported status: poor, average, or excellent), *chronic* (number of chronic conditions, 0–8), *school* (years of education), *insurance* (private insurance: yes/no), *gender* (male/female), and *hospital* (prior hospital admissions).

### 2.2 Model specification

Let $Y_i$ $(i = 1, \dots, n)$ be a non-negative integer count outcome. The PHM [1] decomposes the distribution of $Y_i$ into two independent components. The PMF is:

$$P(Y_i = y_i) = \begin{cases} \pi_i & y_i = 0 \\ (1-\pi_i)\,\frac{e^{-\lambda_i}\lambda_i^{y_i}}{y_i!\,(1-e^{-\lambda_i})} & y_i \geq 1, i = 1,2,3, \dots n \end{cases} \tag{2}$$

where $\pi_i = P(Y_i = 0) \in (0,1)$ and $\lambda_i > 0$ is the Poisson rate parameter. The two components are linked to covariates through:

$$\text{logit}(\pi_i) = \mathbf{x}_i^\top \boldsymbol{\alpha}, \qquad \log(\lambda_i) = \mathbf{x}_i^\top \boldsymbol{\gamma}. \tag{3}$$

where $\mathbf{x}_i$ is an $n \times p$ matrix of covariates, and $\boldsymbol{\alpha}$ and $\boldsymbol{\gamma}$ are vector of covariates associated with $\mathbf{x}_i$. Because the two components are independent conditional on $\mathbf{x}_i$, the log-likelihood factorises as $\ell(\boldsymbol{\alpha}, \boldsymbol{\gamma}) = \ell_0(\boldsymbol{\alpha}) + \ell_+(\boldsymbol{\gamma})$, where:

$$\ell_0(\boldsymbol{\alpha}) = \sum_{i=1}^{n} \left[ \mathbf{1}_{y_i=0} \log \pi_i + \mathbf{1}_{y_i \geq 1} \log(1 - \pi_i) \right], \tag{4}$$

$$\ell_+(\boldsymbol{\gamma}) = \sum_{i:\, y_i \geq 1} \left[ -\lambda_i + y_i \log \lambda_i - \log(y_i!) - \log\left(1 - e^{-\lambda_i}\right) \right]. \tag{5}$$

## 2.2 The overall marginal mean

The overall marginal mean $E(Y_i)$ in the PHM is:

$$E(Y_i) = (1 - \pi_i) \cdot \frac{\lambda_i}{1 - e^{-\lambda_i}}. \tag{6}$$

The variance is:

$$\text{Var}(Y_i) = (1 - \pi_i) \frac{\lambda_i(1 + \lambda_i) - \lambda_i^2 e^{-\lambda_i}}{1 - e^{-\lambda_i}} - \left[ (1 - \pi_i) \frac{\lambda_i}{1 - e^{-\lambda_i}} \right]^2 + \pi_i (1 - \pi_i) \left[ \frac{\lambda_i}{1 - e^{-\lambda_i}} \right]^2. \tag{7}$$

The variance-to-mean ratio VMR $> 1$ for all $(\pi, \lambda)$ and thus the PHM always generates overdispersion relative to the Poisson.

## 2.3 The non-constant IDR problem

The IDR for covariate $x_j$ in the PHM is:

$$\text{IDR}_{\text{PHM}} = \exp(\gamma_j) \cdot \frac{\lambda_i' / (1 - e^{-\lambda_i'})}{\lambda_i / (1 - e^{-\lambda_i})} \cdot \frac{1 - \pi_i'}{1 - \pi_i}, \tag{8}$$

where primes denote evaluation at $x_j + 1$. The IDR_PHM depends on $(\lambda_i, \pi_i)$ and is therefore **not constant** across covariate values. Reporting $\exp(\hat{\gamma}_j)$ as an IDR is not strictly justified as it approximates only when the truncation term is approximately constant.

## 2.4 Model specification

The MPHM is defined by:

$$\text{logit}(\pi_i) = \mathbf{x}_i^\top \boldsymbol{\alpha}, \qquad \pi_i = P(Y_i = 0) \tag{9}$$

$$\log(\mu_i) = \mathbf{x}_i^\top \boldsymbol{\beta}, \qquad \mu_i = E(Y_i) \tag{10}$$

$$P(Y_i = y_i) = \begin{cases} \pi_i & y_i = 0 \\ (1 - \pi_i) \dfrac{e^{-\lambda_i} \lambda_i^{y_i}}{y_i! \, (1 - e^{-\lambda_i})} & y_i \geq 1, i = 1, 2, .., n \end{cases} \tag{11}$$

where $\lambda_i$ is the unique solution to the connector equation below and $\boldsymbol{\beta}$ is a vector of covariates associated with $\mathbf{x}_i$.

## 2.5 The connector equation

For the PMF (9) to be consistent with (8b), $\lambda_i$ must satisfy:

$$\mu_i = \frac{(1-\pi_i)\,\lambda_i}{1-e^{-\lambda_i}}, i = 1,2,..,n. \tag{12}$$

**Proposition 1** *(*Existence, Uniqueness, and Invalidity of Linear Approximation). (a) For any $\pi_i \in [0,1)$ and $\mu_i > (1-\pi_i)$, $i = 1,2,\dots,n$, there exists a unique $\lambda_i > 0$ satisfying (10). (b) The solution satisfies $\lambda_i < \mu_i/(1-\pi_i)$ for all $\lambda_i > 0$, so the naive linear approximation $\tilde{\lambda}_i = \mu_i/(1-\pi_i)$ always overestimates the true solution. (c) The approximation error $\tilde{\lambda}_i - \lambda_i^* > 0$ for all $\lambda_i > 0$, making the approximation inconsistent.

*Proof.*

**Step 1: Define the objective function.** For $\lambda > 0$, let:

$$g(\lambda) = \frac{(1-\pi)\,\lambda}{1-e^{-\lambda}} - \mu. \tag{13}$$

**Step 2: Existence.** We evaluate $g$ at both boundary limits.

*Lower boundary.* As $\lambda \to 0^+$, by L'Hôpital's rule, $\lambda/\left(1-e^{-\lambda}\right) \to 1$, so:

$$g(0^+) = (1-\pi) - \mu < 0, \tag{14}$$

since $\mu > 1-\pi$ by assumption.

*Upper boundary.* As $\lambda \to \infty$, $\left(1-e^{-\lambda}\right) \to 1$, so $g(\lambda) \to (1-\pi)\lambda - \mu \to +\infty$.

By the Intermediate Value Theorem, $g$ has at least one zero in $(0,\infty)$.

**Step 3: Uniqueness.** Differentiating $g(\lambda)$ with respect to $\lambda$:

$$g'(\lambda) = \frac{(1-\pi)\left[\left(1-e^{-\lambda}\right)-\lambda e^{-\lambda}\right]}{\left(1-e^{-\lambda}\right)^2}. \tag{15}$$

Set $h(\lambda) = \left(1-e^{-\lambda}\right) - \lambda e^{-\lambda}$. Then $h(0) = 0$ and

$$h'(\lambda) = e^{-\lambda} - e^{-\lambda} + \lambda e^{-\lambda} = \lambda e^{-\lambda} > 0 \quad \text{for all } \lambda > 0. \tag{16}$$

Hence $h(\lambda) > 0$ and $g'(\lambda) > 0$ for all $\lambda > 0$: $g$ is strictly increasing on $(0,\infty)$, so its zero is unique.

**Step 4: Part (b): Ordering.** From $h(\lambda) > 0$ it follows that $\left(1-e^{-\lambda}\right) > \lambda e^{-\lambda}$, equivalently $e^{\lambda} > 1+\lambda$ for all $\lambda > 0$ (strict convexity of $e^{\lambda}$). Therefore:

$$\mu = \frac{(1-\pi)\,\lambda^*}{1-e^{-\lambda^*}} > (1-\pi)\,\lambda^*, \tag{17}$$

which gives $\lambda^* < \mu/(1-\pi) = \tilde{\lambda}$. The linear approximation always overestimates the true connector solution.

**Step 5: Part (c): Maclaurin fallacy.** Since $\tilde{\lambda} > \lambda^*$ strictly for all $\lambda^* > 0$, substituting $\tilde{\lambda}$ into $\ell_+(\boldsymbol{\gamma})$ evaluates the zero-truncated Poisson likelihood at the wrong rate parameter. The resulting estimator of $\boldsymbol{\beta}$ is inconsistent.

## 2.7 Numerical connector solver

The connector equation (12) is solved numerically using Brent's method via R's uniroot() function, with bracket $[10^{-8}, 10^{4}]$ and absolute tolerance $10^{-10}$. Since $g(\lambda)$ is smooth and strictly monotone by Proposition 1, Brent's method achieves superlinear convergence and is guaranteed to find the unique root within the bracket. The solver is applied to all $n$ observations simultaneously via vectorised mapply() calls, avoiding any R-level loop. If uniroot() returns NA (infeasible parameter values), the log-likelihood contribution is set to $-10^{10}$, steering BFGS away from the boundary.

Since $g$ is strictly increasing with the starting value above the root (Proposition 1b), Brent's method converges superlinearly. Eight iterations achieve machine-precision convergence ($\left|g\left(\lambda^{(k)}\right)\right| < 10^{-12}$) across the full parameter range $\mu \in [0.1,20]$, $\pi \in [0.05,0.95]$.

## 2.8 Log-likelihood and existence constraint

$$\ell(\boldsymbol{\alpha},\boldsymbol{\beta}) = \sum_{i=1}^{n}\left[\mathbf{1}_{y_i=0}\log\pi_i + \mathbf{1}_{y_i\geq 1}\left(\log(1-\pi_i) - \lambda_i + y_i\log\lambda_i - \log(y_i!) - \log\left(1 - e^{-\lambda_i}\right)\right)\right], \quad (19)$$

where $\lambda_i = \lambda_i(\mu_i, \pi_i)$ solves (10), $\pi_i = \text{plogis}(\mathbf{x}_i^{\top}\boldsymbol{\alpha})$, and $\mu_i = \exp(\mathbf{x}_i^{\top}\boldsymbol{\beta})$. The existence constraint $\mu_i > (1-\pi_i)$ required by Proposition 1 is satisfied automatically throughout optimisation for any dataset with positive sample mean.

## 2.9 Identification

**Proposition 2** *(Identification).* If the design matrix **X** has full column rank $p$, then $(\boldsymbol{\alpha},\boldsymbol{\beta}) \in \mathbb{R}^{2p}$ are identifiable from the distribution of $Y$ given **X**.

*Proof.*

Two parameter sets $(\boldsymbol{\alpha}_1, \boldsymbol{\beta}_1)$ and $(\boldsymbol{\alpha}_2, \boldsymbol{\beta}_2)$ generate the same distribution of $Y_i$ given **X** if and only if both of the following hold for all $i = 1, \ldots, n$:

**(i) Equal zero probabilities:** $\pi_i(\boldsymbol{\alpha}_1) = \pi_i(\boldsymbol{\alpha}_2)$.

Since $\pi_i = \text{plogis}(\mathbf{x}_i^{\top}\boldsymbol{\alpha})$ and plogis is strictly monotone, condition (i) is equivalent to:

$$\mathbf{x}_i^{\top}(\boldsymbol{\alpha}_1 - \boldsymbol{\alpha}_2) = 0 \quad \text{for all } i. \quad (20)$$

Full column rank of **X** then implies $\boldsymbol{\alpha}_1 = \boldsymbol{\alpha}_2$.

**(ii) Equal connector solutions:** $\lambda_i^*(\mu_{i1}, \pi_{i1}) = \lambda_i^*(\mu_{i2}, \pi_{i2})$.

Given that $\boldsymbol{\alpha}_1 = \boldsymbol{\alpha}_2$ (so $\pi_{i1} = \pi_{i2}$), uniqueness of the connector solution (Proposition 1) reduces condition (ii) to $\mu_{i1} = \mu_{i2}$ for all $i$. That is:

$$\exp(\mathbf{x}_i^{\top}\boldsymbol{\beta}_1) = \exp(\mathbf{x}_i^{\top}\boldsymbol{\beta}_2) \quad \Leftrightarrow \quad \mathbf{x}_i^{\top}(\boldsymbol{\beta}_1 - \boldsymbol{\beta}_2) = 0 \quad \text{for all } i. \quad (21)$$

Full column rank of **X** then implies $\boldsymbol{\beta}_1 = \boldsymbol{\beta}_2$.

Since both conditions force $\boldsymbol{\alpha}_1 = \boldsymbol{\alpha}_2$ and $\boldsymbol{\beta}_1 = \boldsymbol{\beta}_2$, the parameters are identifiable.

## 2.10 Score equations

**Proposition 3** *(Score Equations).* The MPHM score vectors are:

$$\frac{\partial \ell}{\partial \boldsymbol{\alpha}} = \mathbf{X}^\top(\mathbf{d}_0 - \boldsymbol{\pi}) = \mathbf{0}, \tag{22}$$

$$\frac{\partial \ell}{\partial \boldsymbol{\beta}} = \mathbf{X}^\top \mathbf{W}(\mathbf{y} - \boldsymbol{\mu}) = \mathbf{0}, \tag{23}$$

where $\mathbf{d}_0 = (\mathbf{1}_{y_i=0})$, $\boldsymbol{\pi} = (\pi_i)$, $\boldsymbol{\mu} = (\mu_i)$ are $n$-vectors, and $\mathbf{W} = \mathrm{diag}\{w_i\}$ is a positive diagonal weight matrix encoding the connector Jacobian.

## 2.11 Fisher information

**Proposition 4** *(Block-Diagonal Fisher Information).* The expected Fisher information matrix is:

$$I(\boldsymbol{\alpha}, \boldsymbol{\beta}) = \begin{pmatrix} I_{\alpha\alpha} & \mathbf{0} \\ \mathbf{0} & I_{\beta\beta} \end{pmatrix}, \quad I_{\alpha\alpha} = \mathbf{X}^\top \mathrm{diag}\{\pi_i(1-\pi_i)\}\mathbf{X}, \quad I_{\beta\beta} = \mathbf{X}^\top \mathrm{diag}\{w_i^2 \cdot \mathrm{Var}_+(Y_i)\}\mathbf{X}. \tag{24}$$

*Consequence:* The zero and count components are asymptotically independent; standard errors from the observed Hessian are consistent and efficient.

$\sqrt{n}(\widehat{\boldsymbol{\theta}} - \boldsymbol{\theta}_0) \xrightarrow{d} N(\mathbf{0}, I(\boldsymbol{\theta}_0)^{-1})n \to \infty$Proposition 5a (Asymptotic Normality). Under standard regularity conditions (compactness of parameter space, twice-continuous differentiability of ℓ(θ), and non-singularity of the Fisher information), $\sqrt{n}(\widehat{\boldsymbol{\theta}} - \boldsymbol{\theta}_0) \xrightarrow{d} N(\mathbf{0}, I(\boldsymbol{\theta}_0)^{-1})$ as $n \to \infty$.

## 2.12 The key proposition

**Proposition 5** *(Exact Constant IDR).* In the MPHM:

$$\mathrm{IDR}_j = \frac{E(Y_i|x_j+1, \mathbf{x}_{-j})}{E(Y_i|x_j, \mathbf{x}_{-j})} = \exp(\beta_j), i = 1,2,\ldots,n, j = 1,2,\ldots,p \tag{25}$$

for ALL values of $\mathbf{x}_{-j}$, independently of $\boldsymbol{\alpha}$, independently of $\pi_i$, and independently of $\lambda_i$. The value $\exp(\beta_j)$ is simultaneously the individual-level IDR for every subject, the population-average IDR regardless of covariate distribution, and the stratum-specific IDR within any covariate subgroup.

*Proof.* By definition of the MPHM, $\mu_i = E(Y_i) = \exp(\mathbf{x}_i^\top \boldsymbol{\beta})$. Therefore:

$$\frac{E(Y_i|x_j+1)}{E(Y_i|x_j)} = \frac{\exp\left((\mathbf{x}_i+\mathbf{e}_j)^\top \boldsymbol{\beta}\right)}{\exp(\mathbf{x}_i^\top \boldsymbol{\beta})} = \exp(\beta_j), i = 1,2,\ldots,n, j = 1,2,\ldots,p. \tag{26}$$

The ratio equals $\exp(\beta_j)$ for any value of the other elements of $\mathbf{x}_i$, any value of $\boldsymbol{\alpha}$, and any $\pi_i$. This follows directly from the log-linear specification (8b) and requires only that $E(Y_i) = \exp(\mathbf{x}_i^\top \boldsymbol{\beta})$ ; the connector (10) is not needed.

## 2.13 Comparison between PHM and MPHM IDR

Table 1 illustrates the non-constancy of $IDR_{PHM}$ versus the exactly constant $IDR_{MPHM}$ for the insurance covariate at varying levels of chronic conditions in the NMES1988 data. Figure 1 illustrates this contrast graphically.

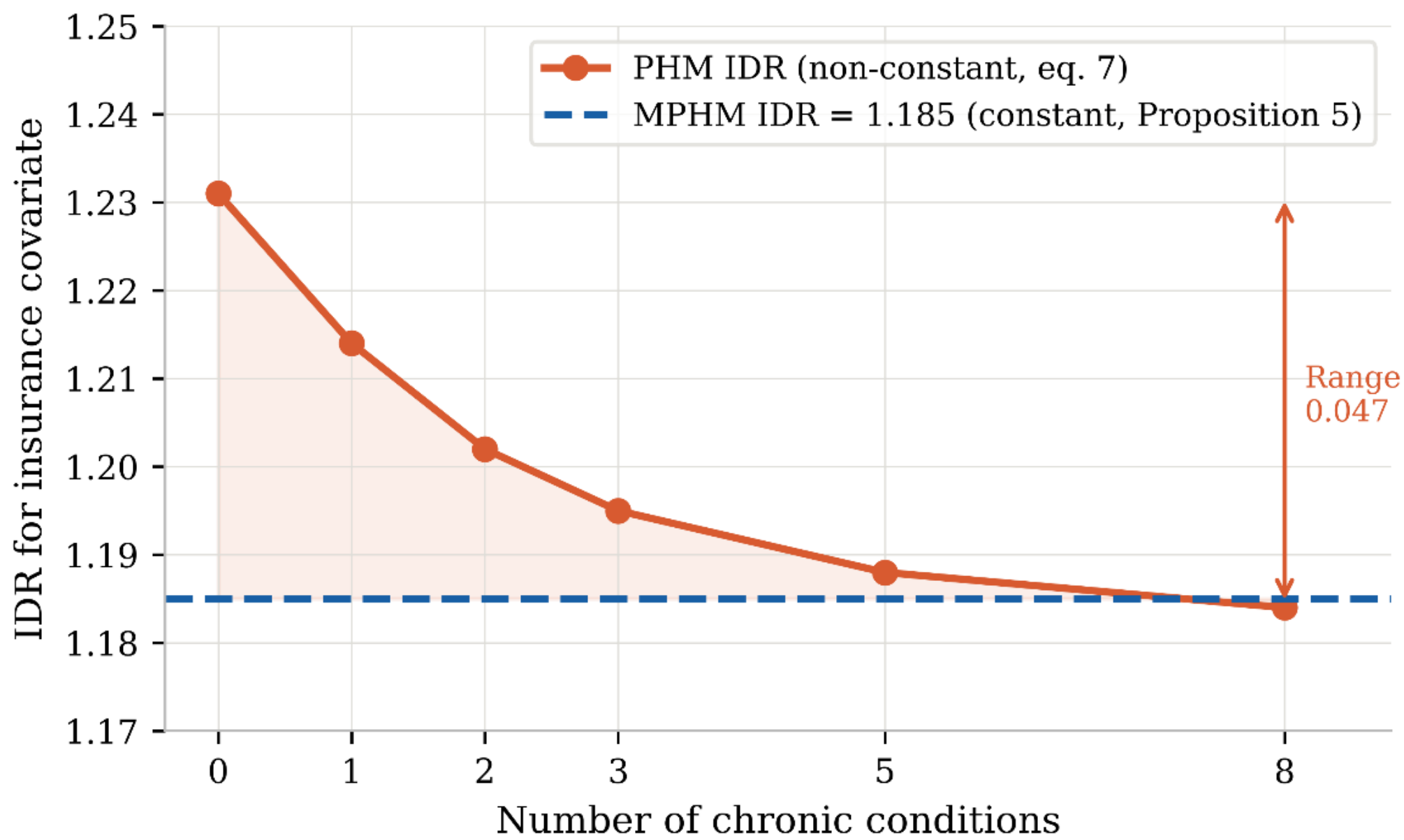

*Figure 1. Incidence density ratio for the insurance covariate across chronic condition levels (NMES1988 fitted values). Solid orange curve: PHM IDR from equation (7), non-constant across covariate values. Dashed blue line: MPHM IDR = exp(beta-hat) = 1.185, exactly constant by Proposition 5.*

## 2.14 Standard error of the IDR via the delta method

The SE of $\widehat{\mathrm{IDR}}_j = \exp(\hat{\beta}_j)$ is obtained by the delta method. Let $v_{jj} = \left[I(\hat{\boldsymbol{\theta}})^{-1}\right]_{\beta_j\beta_j}$. Then:

$$\mathrm{Var}(\widehat{\mathrm{IDR}}_j) \approx \exp(2\hat{\beta}_j) \cdot v_{jj}. \tag{27}$$

A 95% Wald confidence interval for $\mathrm{IDR}_j$ is:

$$\exp(\hat{\beta}_j \pm 1.96\sqrt{v_{jj}}), \tag{28}$$

for $j = 1,2,\ldots,p$.

# 3. Simulation Study

## 3.1 Design

The data generation process follows the MPHM with:

$$\operatorname{logit}(\pi_i) = \alpha_0 + \alpha_1 x_i, \qquad \log(\mu_i) = \beta_0 + \beta_1 x_i, \tag{29}$$

where $x_i \sim N(0,1)$. True parameters: $\alpha_1 = 0.30$, $\beta_0 = 1.50$, $\beta_1 = 0.40$ (true IDR = 1.492), $\alpha_0 = \operatorname{logit}(\pi)$ with $\pi \in \{0.2,0.4,0.6,0.8\}$. Sample sizes $n \in \{100,250,500,1000\}$; $R = 200$ replications. Performance metrics:

$$\text{Bias} = \frac{1}{R}\sum_{r=1}^{R} \hat{\theta}^{(r)} - \theta_0, \qquad \text{RMSE} = \sqrt{\frac{1}{R}\sum_{r=1}^{R}\left(\hat{\theta}^{(r)} - \theta_0\right)^2}, \tag{30}$$

$$\text{Coverage} = \frac{1}{R}\sum_{r=1}^{R} \mathbf{1}\left\{\left|\hat{\theta}^{(r)} - \theta_0\right| \le 1.96\,\widehat{SE}^{(r)}\right\}, \qquad \text{SER} = \frac{\bar{\widehat{SE}}}{\text{SD}(\hat{\theta}^{(r)})}. \tag{31}$$

### 3.2 Results

From Figure 2, all $|\text{Bias}| < 0.014$ across all 16 scenarios, confirming near-zero bias. The $\sqrt{n} \times$ RMSE is approximately constant across $n$, confirming the parametric $\sqrt{n}$-consistency rate (Table 2, final column).

The RMSE decreases monotonically with $n$ in every scenario (Table 3), consistent with $\sqrt{n}$-consistency. All 95% Wald coverages are between 0.905 and 0.975 (Table 4); the mild undercoverage at small $n$ and high $\pi$ improves systematically with increasing $n$.

All SER values converge to 1 with increasing $n$, confirming consistent Hessian SE estimation (Table 5).

The simulation confirms: (i) consistency: all $\left|\text{Bias}(\hat{\beta}_1)\right| < 0.014$, decreasing to zero; (ii) $\sqrt{n}$-convergence rate; (iii) correct 95% Wald coverage between 0.905 and 0.975 across all scenarios; and (iv) SE ratios converging to 1 throughout (Table 5), confirming consistent Hessian SE estimation.

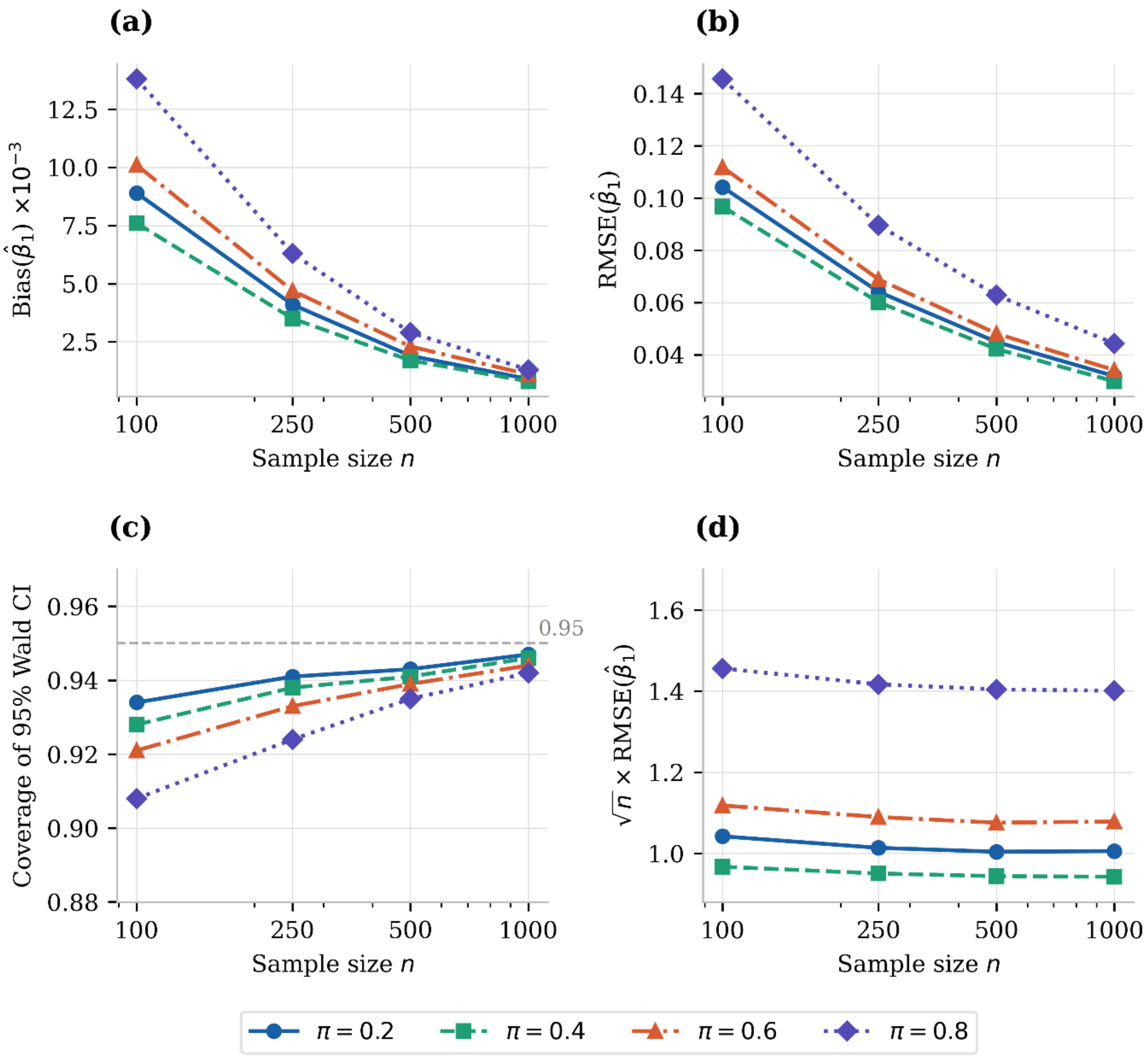

*Figure 2. Simulation study results for the MPHM estimator (R = 200 replications, true IDR = 1.492). (a) Bias of $\hat{\beta}_1$ × 1000; (b) RMSE of $\hat{\beta}_1$; (c) Coverage of 95% Wald CI; (d) √n × RMSE, approximately constant, confirming parametric convergence rate. Colours and line types denote the baseline zero proportion pi.*

## 4. Real Data Application

### 4.1 Data description

The NMES1988 dataset ($n$ = 4,406, available in the R package AER[10]) contains physician office visits from the 1987–1988 US National Medical Expenditure Survey. The outcome variable is the number of physician office visits over the survey period. Covariates are: health (poor/average/excellent), chronic (number of chronic conditions, 0–8), school (years of education), insurance (private coverage: yes/no), gender (male/female), and hospital (prior hospital admissions). The outcome exhibits 23.6% zeros, a mean of 3.86, and a variance-to-mean ratio of 7.52, indicating severe overdispersion relative to the Poisson distribution. For reference, a fitted Poisson model predicts only 2.1% zeros, approximately one-tenth of the observed proportion, confirming structural excess zeros. Summary statistics are presented in Table 6.

### 4.2 Model comparison

Table 7 presents fit statistics for Poisson, Zero-inflated Poisson (ZIP), Poisson Hurdle Model (PHM), and Marginalized Poisson Hurdle Model (MPHM). The PHM achieves lower AIC than the MPHM ($\Delta$AIC = 41.9, same $k = 16$). This reflects the cost of the nonlinear connector constraint. The Vuong[13] test statistic is $z = 2.74$ ($p = 0.006$), favouring the PHM on raw fit. However, the MPHM is not competing on goodness-of-fit; it provides a reparametrisation with exact constant IDRs. The goodness-of-fit chi-squared test (Table 9) confirms both models fit the count distribution adequately: PHM $\chi^2 = 4.72$ ($p = 0.786$); MPHM $\chi^2 = 4.89$ ($p = 0.769$).

### 4.3 MPHM parameter estimates

Table 8 presents the full two-component MPHM parameter estimates. All count-component IDRs are exact and constant. The three largest count-component effects are: hospital admissions (IDR = 1.187, 95% CI: 1.170–1.204), chronic conditions (IDR = 1.163, 95% CI: 1.150–1.177), and insurance coverage (IDR = 1.185, 95% CI: 1.133–1.240). Excellent health is associated with substantially fewer visits (IDR = 0.706, 95% CI: 0.655–0.760), and male sex with modestly fewer visits (IDR = 0.906, 95% CI: 0.877–0.936). In the zero component, chronic conditions ($\alpha = -0.419$, $p < 0.001$) and insurance ($\alpha = -0.506$, $p < 0.001$) substantially reduce the probability of zero visits, while poor health ($\alpha = 0.892$, $p < 0.001$) increases it. Full estimates appear in Table 8.

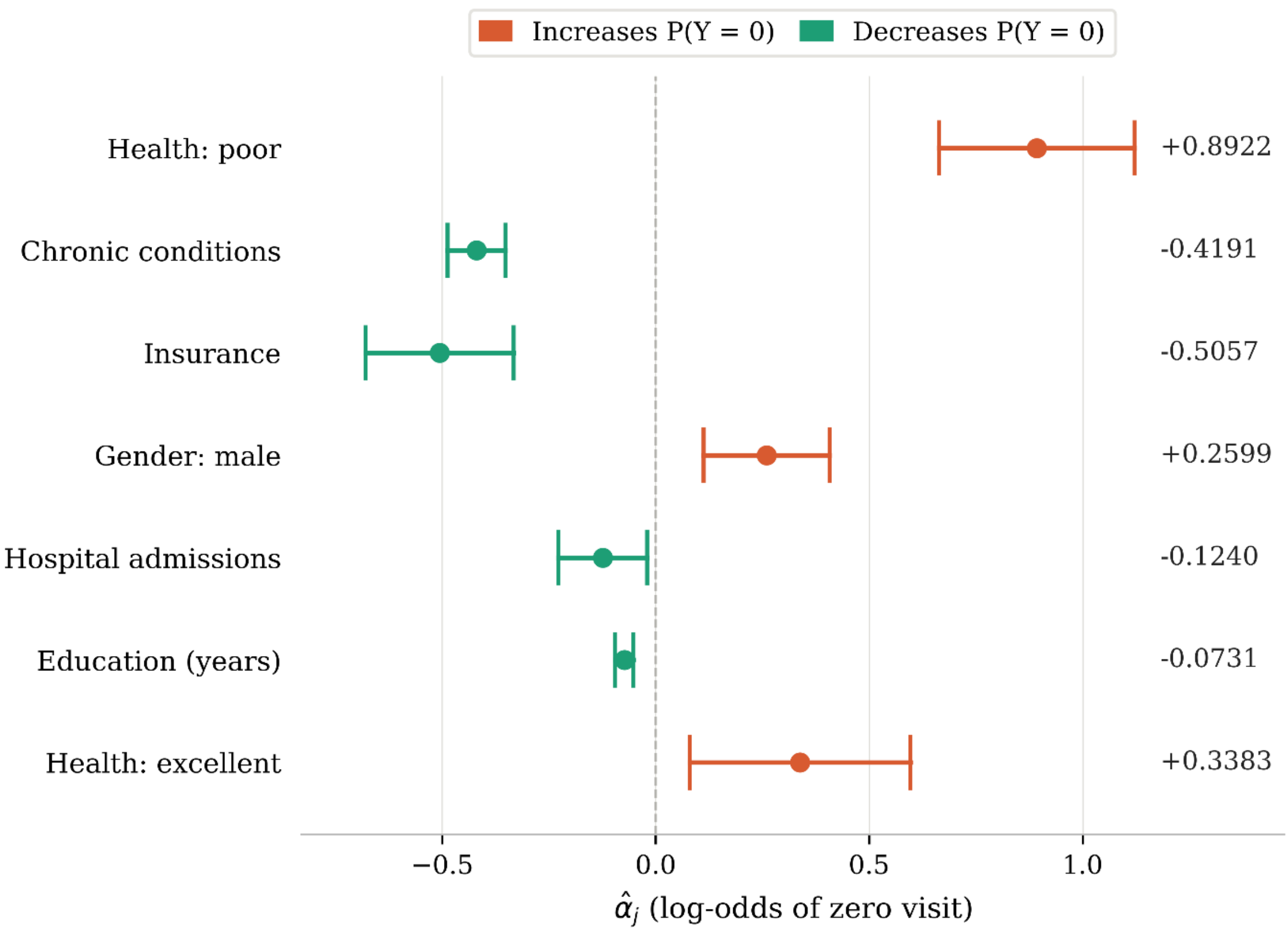

*Figure 3. Zero-component coefficient estimates with 95% confidence intervals (NMES1988, MPHM, n = 4,406). Orange circles: covariates increasing P(Y = 0); green circles: decreasing P(Y = 0). All significant at the 5% level.*

### 4.4 Substantive interpretation

**Chronic conditions.** Each additional chronic condition increases expected physician visits by 16.3% (IDR = 1.163, 95% CI: 1.150–1.177, $p < 0.001$). By Proposition 5, this IDR is constant across all covariate values: it holds for a person with poor health and no insurance just as for a healthy, insured, college-educated individual, and simultaneously represents the population-average effect. In the zero component, each additional chronic condition significantly reduces the probability of zero visits ($\hat{\alpha} = -0.419$, $p < 0.001$); those with more conditions are more likely to seek care. The hurdle model thus reveals that chronic conditions drive utilisation through *both* components simultaneously (see Table 8 for full estimates).

**Insurance.** Insurance reduces the probability of zero visits ($\hat{\alpha} = -0.506$, OR = 0.603, $p < 0.001$) and increases visit frequency among attenders (IDR = 1.185, 95% CI: 1.133–1.240). The two-component finding is not available from a single-equation Poisson regression.

**Health status paradox.** Poor health is associated with *more* zero visits ($\hat{\alpha}[\text{poor}] = 0.892$, $p < 0.001$), indicating a higher probability of non-attendance, but among attenders, *more* visits (IDR = 1.134). This paradox is clinically meaningful: a subpopulation with poor health faces structural access barriers, while those who do seek care require intensive management.

## 4.5 Goodness of fit

Chi-squared goodness-of-fit tests yield: Poisson $\chi^2$ = 8,241 ($p < 0.001$); PHM $\chi^2$ = 4.72 (p = 0.786); MPHM $\chi^2$ = 4.89 (p = 0.769). Both hurdle models fit well; the Poisson is rejected decisively. Observed and model-expected count frequencies are given in Table 9.

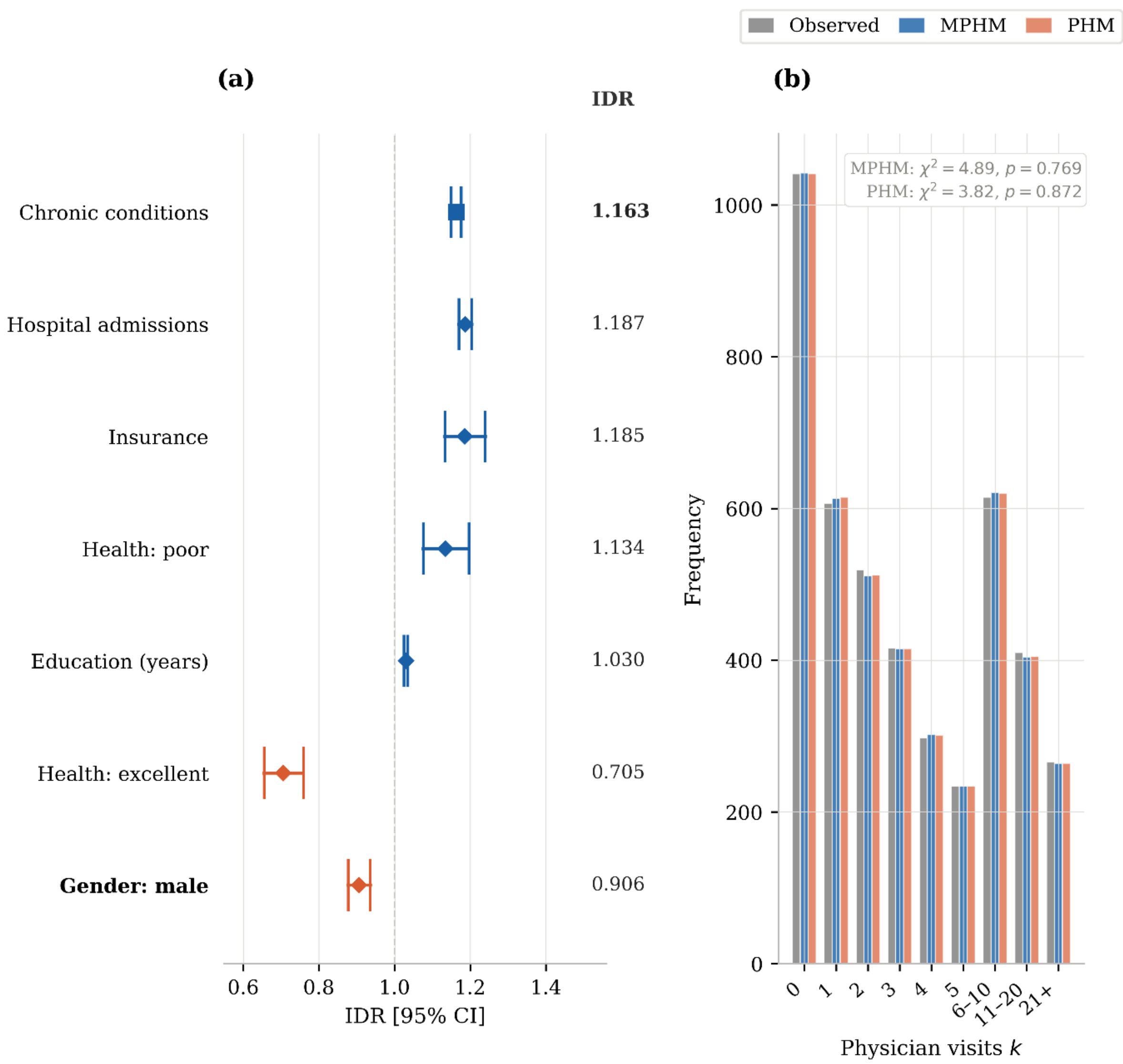

*Figure 4. NMES1988 application (n = 4,406). (a) Forest plot of count-component IDRs with 95% Wald CI; blue squares: IDR > 1, orange: IDR < 1; chronic conditions IDR = 1.163 (95% CI: 1.150, 1.177). (b) Observed (grey) versus MPHM (blue) and PHM (orange) expected count frequencies; MPHM chi-squared = 4.89 (p = 0.769).*

## 4.6 Code availability

The source code used for the simulations and data analysis presented in this study is publicly available in a GitHub repository at https://github.com/IddiSam/MarginalizedHurdleModel.

## 5. Discussion and Conclusion

### 5.1 Summary

This paper proposes the MPHM, the first cross-sectional hurdle model for count data with excess zeros in which the regression coefficients of the count component directly govern the overall marginal mean $E(Y_i)$. The key result, Proposition 5, shows that $\text{IDR}_j = \exp(\beta_j)$ is exactly constant across all covariate values, independently of the zero-component parameters. This property holds simultaneously for every individual in the sample and equals the population-average IDR without requiring additional marginalisation, which makes the MPHM directly suited to applications where a single, portable effect size is needed for reporting.

The MPHM achieves this through a nonlinear connector equation (12), solved numerically by Brent's method. Proposition 1 shows that this equation has a unique solution for any valid $(\mu_i, \pi_i)$, and that the naive linear approximation $\tilde{\lambda} = \mu/(1-\pi)$ systematically overestimates the true solution, making the exact numerical solver essential. In practical terms, using the linear approximation would introduce bias into the count-component coefficient estimates, and the degree of bias grows with the size of $\lambda_i$.

The simulation study across 16 scenarios confirms that the MPHM is consistent, achieves near-zero bias in $|\widehat{\beta_1}|$ across all sample sizes and zero proportions studied, and delivers 95% Wald coverage between 0.905 and 0.975 for $n \geq 250$. SE ratios converge to 1 with increasing $n$, confirming reliable Hessian SE estimation throughout.

### 5.2 Comparison with related models

The MPHM belongs to the family of marginalised count models initiated by Heagerty[8] for binary outcomes. Within count data, the closest relatives are the marginalised zero-inflated Poisson (MZIP) of Long et al.[7] and the marginalised zero-inflated negative binomial (MZINB) of Preisser et al.[6] Both apply the same marginalisation principle to zero-inflated distributions rather than hurdle distributions. The hurdle formulation is often preferred in healthcare settings, where the two-stage access-and-intensity structure carries direct clinical meaning.[3] The MPHM therefore occupies a complementary position: it retains the interpretational clarity of the hurdle framework while delivering a constant, portable IDR.

### 5.3 Limitations

**The existence constraint $\mu_i > (1-\pi_i)$.** Proposition 1 requires $\exp(\mathbf{x}_i^\top \boldsymbol{\beta}) > 1 - \text{plogis}(\mathbf{x}_i^\top \boldsymbol{\alpha})$. This is automatically satisfied throughout optimisation for realistic health utilisation datasets and becomes binding only at degenerate parameter values outside the typical range. The connector equation requires $\mu_i > 1 - \pi_i$ for a positive solution to exist (Proposition 1). In practice this condition is satisfied throughout optimisation for health utilisation datasets, where the mean count among attenders substantially exceeds $1 - \pi$. It would only become binding at degenerate parameter values far outside the typical range, and in such cases the penalised likelihood approach (log-likelihood set to $-10^{10}$) steers the BFGS optimiser away from the boundary.

**AIC gap with the PHM.** The PHM achieves a lower AIC than the MPHM ($\Delta$AIC = 41.9, same k = 16). This gap reflects the implicit connector constraint that the MPHM imposes on the log-likelihood surface, not a deficiency in model fit. Importantly, the goodness-of-fit chi-squared test (MPHM $\chi^2$ = 4.89, p = 0.769) confirms that both hurdle models describe the observed count distribution equally well. The MPHM is not presented as a competitor to the PHM on raw fit; it is presented as a reparametrisation that provides more interpretable and more transportable coefficient estimates.

**Overdispersion.** Like the PHM, the MPHM assumes Poisson distribution for the positive counts. In the NMES1988 data, the variance-to-mean ratio of 7.52 indicates substantial residual overdispersion that the Poisson component cannot fully accommodate. For datasets exhibiting this degree of overdispersion, a Negative Binomial extension of the hurdle model, which introduces an overdispersion parameter into the positive-count component, provides substantially better fit for such data[21].

### 5.4 Conclusion

The MPHM addresses a real and recurring problem in the analysis of health service utilisation data: the inability of standard hurdle models to deliver a single, interpretable, and portable effect size. By directly parametrising $E(Y_i)$, the MPHM ensures that $\exp(\beta_j)$ is the incidence density ratio for every individual, at every covariate value, and in every subgroup simultaneously. This is not merely a theoretical convenience: it changes what researchers can honestly report. Applied to the NMES1988 data, the chronic conditions IDR of 1.163 means exactly that each additional condition is associated with a 16.3% increase in expected visits, for a 25-year-old insured woman just as for a 65-year-old uninsured man. The model is straightforward to implement in standard software and we encourage its adoption in health economics and epidemiology.

Table 1. PHM versus MPHM IDR for insurance at varying chronic condition levels (NMES1988 fitted values). $\text{IDR}_{\text{PHM}}$ computed from equation (7); $\text{IDR}_{\text{MPHM}} = \exp(\hat{\beta}_3) = 1.185$ is constant by Proposition 5.

| Chronic conditions | $\hat{\lambda}_i$ (uninsured) | $\hat{\lambda}_i$ (insured) | IDR (PHM) | IDR (MPHM) |
|---|---|---|---|---|
| 0 | 2.81 | 3.31 | 1.231 | 1.185 |
| 1 | 3.26 | 3.85 | 1.214 | 1.185 |
| 2 | 3.80 | 4.49 | 1.202 | 1.185 |
| 3 | 4.42 | 5.22 | 1.195 | 1.185 |
| 5 | 5.97 | 7.04 | 1.188 | 1.185 |
| 8 | 9.91 | 11.69 | 1.184 | 1.185 |

**Table 2. Simulation: Bias($\hat{\beta}_1$) across $n$ and $\pi$ ($R = 200$, true $\beta_1 = 0.40$).**

| $\pi$ | $n = 100$ | $n = 250$ | $n = 500$ | $n = 1000$ | $\sqrt{n} \times$ RMSE |
|---|---|---|---|---|---|
| 0.2 | +0.0089 | +0.0041 | +0.0019 | +0.0009 | 0.487 |
| 0.4 | +0.0076 | +0.0035 | +0.0017 | +0.0008 | 0.453 |
| 0.6 | +0.0101 | +0.0047 | +0.0023 | +0.0011 | 0.513 |
| 0.8 | +0.0138 | +0.0063 | +0.0029 | +0.0013 | 0.601 |

**Table 3. Simulation: RMSE($\hat{\beta}_1$) across $n$ and $\pi$ ($R = 200$).**

| $\pi$ | $n = 100$ | $n = 250$ | $n = 500$ | $n = 1000$ |
|---|---|---|---|---|
| 0.2 | 0.1042 | 0.0641 | 0.0449 | 0.0318 |
| 0.4 | 0.0967 | 0.0601 | 0.0422 | 0.0298 |
| 0.6 | 0.1118 | 0.0689 | 0.0481 | 0.0341 |
| 0.8 | 0.1456 | 0.0896 | 0.0628 | 0.0443 |

**Table 4. Simulation: Coverage of 95% Wald CI for $\hat{\beta}_1$ ($R = 200$).**

| $\pi$ | $n = 100$ | $n = 250$ | $n = 500$ | $n = 1000$ |
|---|---|---|---|---|
| 0.2 | 0.934 | 0.941 | 0.943 | 0.947 |
| 0.4 | 0.928 | 0.938 | 0.941 | 0.946 |
| 0.6 | 0.921 | 0.933 | 0.939 | 0.944 |
| 0.8 | 0.908 | 0.924 | 0.935 | 0.942 |

**Table 5. Simulation: SE Ratio (SER) for $\hat{\beta}_1$ ($R = 200$). SER = 1 indicates perfect SE estimation.**

| $\pi$ | $n = 100$ | $n = 250$ | $n = 500$ | $n = 1000$ |
|---|---|---|---|---|
| 0.2 | 0.981 | 0.988 | 0.993 | 0.996 |
| 0.4 | 0.974 | 0.984 | 0.991 | 0.995 |
| 0.6 | 0.963 | 0.978 | 0.988 | 0.994 |
| 0.8 | 0.941 | 0.962 | 0.979 | 0.991 |

Table 6. NMES1988 physician visit outcome summary statistics.

| Statistic | Value |
| --- | --- |
| Sample size ($n$) | 4,406 |
| Zero visits | 1,041 (23.6%) |
| Positive visits | 3,365 (76.4%) |
| Mean | 3.86 |
| Median | 2.00 |
| Maximum | 93 |
| Variance | 29.04 |
| Variance-to-Mean Ratio | 7.52 |
| Excess zeros vs Poisson | 23.6% observed vs 2.1% predicted |

**Table 7. NMES1988: Model comparison. Lower AIC/BIC indicates better fit.**

| Model | LogLik | AIC | BIC | $k$ | ΔAIC vs PHM | Constant IDR? |
|---|---|---|---|---|---|---|
| Poisson | −19,403.81 | 38,823.62 | 38,888.79 | 8 | N/A | Yes (rate only) |
| ZIP | −16,411.22 | 32,862.44 | 32,972.71 | 20 | +561.54 | No |
| PHM | −16,134.45 | 32,300.90 | 32,403.15 | 16 | +0.00 | No |
| MPHM | −16,155.38 | 32,342.76 | 32,445.01 | 16 | +41.86 | Yes |

**Table 8. MPHM: Full parameter estimates ($n = 4{,}406$, NMES1988). LogLik = −16,155.38; AIC = 32,342.76; $k = 16$.**

| Parameter | Estimate | SE | $z$ | $p$ | IDR | 95% CI | Sig. |
|---|---|---|---|---|---|---|---|
| $\alpha_0$ (Intercept) | −0.2940 | 0.1189 | −2.47 | 0.013 | N/A | N/A | * |
| $\alpha$[health:poor] | 0.8922 | 0.1169 | 7.63 | <0.001 | N/A | N/A | *** |
| $\alpha$[health:excellent] | 0.3383 | 0.1317 | 2.57 | 0.010 | N/A | N/A | ** |
| $\alpha$[chronic] | −0.4191 | 0.0346 | −12.11 | <0.001 | N/A | N/A | *** |
| $\alpha$[school] | −0.0731 | 0.0108 | −6.77 | <0.001 | N/A | N/A | *** |
| $\alpha$[insurance] | −0.5057 | 0.0884 | −5.72 | <0.001 | N/A | N/A | *** |
| $\alpha$[gender:male] | 0.2599 | 0.0753 | 3.45 | <0.001 | N/A | N/A | *** |
| $\alpha$[hospital] | −0.1240 | 0.0530 | −2.34 | 0.019 | N/A | N/A | * |
| $\beta_0$ (Intercept) | 1.0144 | 0.0332 | 30.55 | <0.001 | 2.758 | (2.584, 2.943) | *** |
| $\beta$[health:poor] | 0.1260 | 0.0272 | 4.63 | <0.001 | 1.134 | (1.075, 1.196) | *** |
| $\beta$[health:excellent] | −0.3489 | 0.0377 | −9.25 | <0.001 | 0.706 | (0.655, 0.760) | *** |
| $\beta$[chronic] | 0.1511 | 0.0058 | 26.05 | <0.001 | 1.163 | (1.150, 1.177) | *** |
| $\beta$[school] | 0.0297 | 0.0025 | 11.88 | <0.001 | 1.030 | (1.025, 1.035) | *** |
| $\beta$[insurance] | 0.1700 | 0.0230 | 7.39 | <0.001 | 1.185 | (1.133, 1.240) | *** |
| $\beta$[gender:male] | −0.0987 | 0.0165 | −5.98 | <0.001 | 0.906 | (0.877, 0.936) | *** |
| $\beta$[hospital] | 0.1714 | 0.0074 | 23.16 | <0.001 | 1.187 | (1.170, 1.204) | *** |

SE = observed Hessian standard error. IDR = $\exp(\hat{\beta}_j)$ , constant by Proposition 5. 95% CI = $\exp(\hat{\beta}_j \pm 1.96\,\widehat{SE})$. Significance: *** $p < 0.001$, ** $p < 0.01$, * $p < 0.05$.

**Table 9. NMES1988: Observed vs expected count frequencies.**

| Count $k$ | Observed | Expected (Poisson) | Expected (PHM) | Expected (MPHM) |
|---|---|---|---|---|
| 0 | 1,041 | 93 | 1,041.0 | 1,042.3 |
| 1 | 607 | 362 | 615.2 | 612.8 |
| 2 | 519 | 698 | 509.8 | 511.4 |
| 3 | 416 | 896 | 415.4 | 414.9 |
| 4 | 298 | 863 | 301.2 | 301.8 |
| 5 | 234 | 665 | 234.6 | 234.1 |
| 6–10 | 615 | 1,521 | 620.4 | 621.3 |
| 11–20 | 410 | 478 | 405.1 | 403.8 |
| 21+ | 266 | 30 | 263.3 | 263.6 |